\documentclass[twocolumn,twoside]{article}
\usepackage{graphics}
\newcommand{\h}{\linebreak \hspace*{3ex}}
\newcommand{\hb}{\\ \hspace*{2ex}}

\begin{document}
\title{STAR FORMATION HISTORY IN THE GALACTIC THIN DISK}
\author{V.A.\,Marsakov$^{1}$, M.V.\,Shapovalov$^{1}$, T.V.\,Borkova$^{1}$\\[2mm] 
\begin{tabular}{l}
 $^1$ Southern Federal University\hb
 Rostov-on-Don 344090 Russia, \\
 {\em marsakov@ip.rsu.ru}\\
\end{tabular}
}
\date{}
\maketitle

ABSTRACT.
We analyze the relations between the relative magnesium abundances in 
stars, [Mg/Fe], and their metallicities, Galactic orbital elements, and 
ages. The relative magnesium abundances
in metal-poor thin-disk stars have been found to systematically 
decrease with increasing stellar orbital radii.
This behavior suggests that, first, the 
star formation rate decreases with increasing Galactocentric distance and,
second, there was no star formation for some time outside the solar
circle while this process was continuous within the solar circle. The 
decrease in the star formation rate with increasing Galactocentric 
distance is responsible for the existence of a negative radial 
metallicity gradient (grad$_{R}$[Fe/H]=($-0.05 \pm 0.01$)~kpc$^{-1}$) in 
the disk. At the same time the relative magnesium abundance exhibits no 
radial gradient. We discovered that in the thin disk there is not 
only the connection between age and metallicity, but between  age 
and relative magnesium abundance also. 
It is in detail considered the influence of selective effects on the 
form of both age~-- metallicity and age~-- relative magnesium abundance 
diagrams. It is shown that the first several billion years of the 
formation of the thin disk interstellar medium in it was on the 
average sufficiently rich in heavy elements. At the same time the 
relative magnesium abundance exhibits no 
radial gradient.
We discovered that in the thin disk there is not only the connection 
between age and metallicity, but between  age and relative magnesium 
abundance also. It is in detail considered the influence of selective 
effects on the form of both age~-- metallicity and age~-- relative 
magnesium abundance diagrams. It is shown that the first several 
billion years of the formation of the thin disk interstellar medium 
in it was on average sufficiently rich in heavy elements 
($\langle$[Fe/H]$\rangle \approx -0.22)$, badly mixed 
($\sigma_{\textrm{[Fe/H]}} \approx 0.21)$, and the average relative 
magnesium abundance was comparatively high 
$(\langle$[Mg/Fe]$\rangle \approx 0.10)$. Approximately 5~billion 
years ago average metallicity began to systematically increase, 
and its dispersion and the average relative magnesium abundance~-- 
to decrease. These properties may be explained by an increase in 
star formation rate with the simultaneous intensification of the 
processes of mixing the interstellar medium in the thin disk, 
provoke possible by interaction the Galaxy with the completely 
massive galaxy-satellite.
\\[1mm]

{\bf Galaxy (Milky Way), stellar chemical composition, thin disk, 
Galactic evolution.}\\[2mm]


The chemical composition of low-mass 
main-sequence stars can be used to estimate the star 
formation rate and as the time scale of a chemically evolving 
closed system. Thus, for example, the $\alpha$-elements (O, 
Mg, Si, S, Ca and Ti) together with a small number of iron 
atoms are currently believed to be synthesized in the high-mass 
($M>10M_\odot$) asymptotic-giant-branch 
progenitors of type\,II supernovae, while the 
bulk of the iron-group elements are produced during type\,Ia 
supernova explosions. Beginning from the paper by Tinsley~(1979),
the negative trend in the [$\alpha$/Fe] ratio as a function of metallicity
observed in the Galaxy has been assumed to be due to a difference in 
the evolution times of these stars. Indeed, the evolution time scale for 
type\,II supernovae is only $\approx 30$~Myr while low massive SNe\,Ia
explosions begin only in $\approx (0.5\div 1.5$)~Gyr. The higher the 
star formation rate in the system, the larger the metallicity at 
which the knee attributable to the onset of SNe\,Ia explosions 
which result in an enrichment of the interstellar medium with 
iron-group elements, will be observed in 
the [$\alpha$/Fe]--[Fe/H] relation. The lower the star formation rate in 
the system, the steeper the further decrease in the [$\alpha$/Fe] ratio 
with increasing total metallicity. If star formation in the system is
halted altogether then the source of $\alpha$-elements (i.\,e., SNe\,II)
will vanish and only SNe\,Ia will enrich the interstellar medium with 
iron-group elements; therefore the [$\alpha$/Fe] ratio will decrease 
suddenly.

Since more than 90\,\% of the stars in the immediate solar 
neighborhood belong to the youngest (in the Galaxy) thin-disk 
subsystem, the chemical composition of this subsystem has been 
studied in greatest detail. However the sizes of the original 
samples in all works were very limited; this is probably the 
reason why the results and conclusions are occasionally 
in conflict with one another. 

Since the published results disagree, analyzing the relations 
between the relative abundances of $\alpha$-elements and metallicity 
and other parameters of thin-disk stars based on a much larger 
statistical material seems very topical. At the beginning of this 
work we analyze the chemical properties of thin-disk stars using
data from our compiled catalog of spectroscopically determined magnesium 
abundances (Borkova and Marsakov~2005). 
Almost all of the published magnesium abundances in dwarfs
and subgiants in the solar neighborhood determined by synthetic 
modeling of high-dispersion spectra as of December~2003 were gathered in 
our catalog. This catalog is several times larger than any homogeneous
sample that has been used until now to analyze the chemical evolution of 
the Galaxy.

The relative magnesium abundances in the catalog were derived for 
867~stars using a three-pass iterative averaging procedure with a 
weight assigned to each primary source and each individual determination. 

Since our main goal is to analyze the relations between the chemical 
composition and other parameters of thin-disk stars, we identified the 
latter solely according to kinematical criteria. 
The technique for identified the stars for which the probabilities 
of belonging to the thin disk is higher than the probability 
of belonging to the thick disk based on the dispersions of the space 
velocity components and the mean 
rotational velocity of both subsystem at the solar Galactocentric 
distance was taken from Bensby et al.~(2003).

\begin{figure}
\resizebox{\hsize}{!}
{\includegraphics{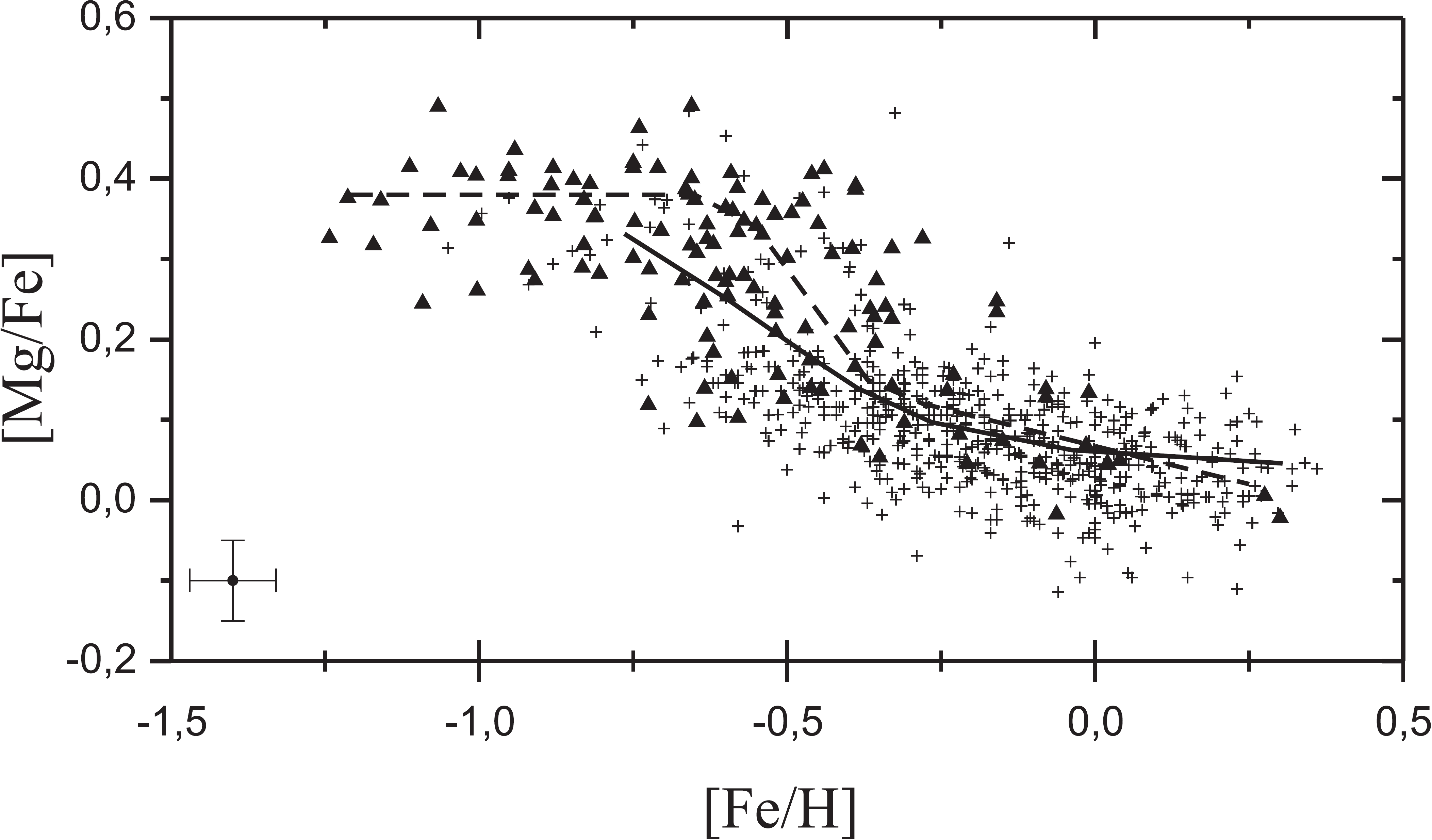}}
\caption{Relation between metallicity and relative magnesium 
abundance for the stars of both disk subsystems: the crosses and 
triangles indicate the thin-disk and thick-disk stars, respectively. 
The broken curves represent the median lines of the relations for 
the thin (solid line) and thick (dotted line) disks drawn by eye 
halfway between the upper and lower envelopes. The error bars are shown}
\label{fig_1}
\end{figure}

Figure\,\ref{fig_1} shows the [Fe/H]--[Mg/Fe] diagram where the 
thin-disk and thick-disk stars are denoted by different symbols. 
We see that the sequence for the thin disk in the metallicity range 
($-1.0 <[Fe/H]<-0.4$) lies systematically lower than that for the 
thick disk. This suggests that the bulk of 
the thick-disk stars formed long before the onset of massive star
formation in the thin disk. The [Mg/Fe] ratio begins to 
decrease with increasing metallicity in the thin disk immediately 
after the formation of the first stars in it, i.\,e., from 
$[Fe/H]\approx -1.0$. This is considerably farther to the left 
in the diagram than in the thick disk where the point of a sharp
decrease is observed at $[Fe/H]\approx 0.5$. Hence, in the 
metal-poor interstellar matter from which the first thin-disk stars 
subsequently began to form, the enrichment with SNe\,II ejecta was 
less intense before this. It is quite probable that such metal-poor 
matter with a high relative magnesium abundance came into the thin 
disk as a result of accretion from regions with a different history 
of chemical evolution. The fast decrease in the relative magnesium 
abundance in the thin disk as the metallicity increases from 
$\approx -1.0$~dex to $\approx -0.7$ suggests that the star 
formation rate in it was initially low but it then suddenly increased,
which subsequently led to a attending of the [Fe/H]--[Mg/Fe] relation.
Subsequently when passing to stars with metallicities higher than 
the solar value, the slope of the [Mg/Fe]--[Fe/H] relation virtually 
vanishes, which is indicative of a new increase in the star formation
rate and stabilization of the ratio of the 
contributions from supernovae (SN\,II/SN\,Ia) to the enrichment of 
the interstellar medium in the thin disk since then.

\begin{figure}
\resizebox{\hsize}{!}
{\includegraphics{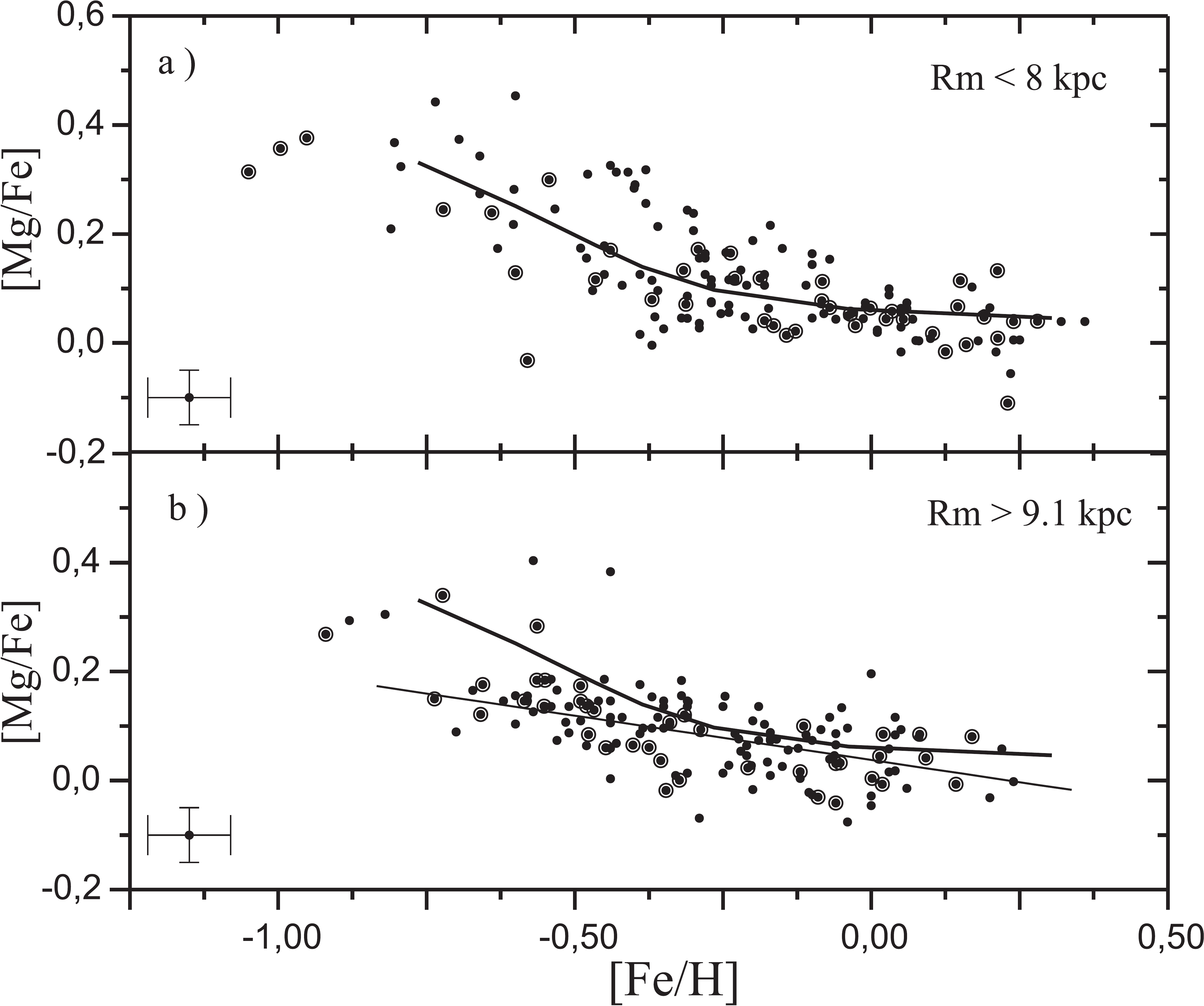}}
\caption{Relation between metallicity and relative magnesium 
abundance for the thin-disk stars in two ranges in
mean orbital radii $R_m < 8$,  and $>9.1$~kpc. The broken 
curves in all panels represent the median curve for the thin disk.
Thin line in panel (b)~-- straight regression for stars with 
$[Mg/Fe] < 0.2$. Is evident the decrease of the relative abundance 
of magnesium in the metal poor ($[Fe/H]<-0.4$) stars with an 
increase in the radii of the orbits.}
\label{fig_2}
\end{figure}

Let us now verify whether the 
positions of the thin-disk stars in [Mg/Fe] the [Fe/H]--[Mg/Fe] 
diagram depend on their mean orbital radii? Figure\,\ref{fig_2} 
shows two diagrams for the thin-disk stars with low ($R_m < 8$~kpc) 
and large ($R_m > 9.1$~kpc) orbits. Our median 
sequence is plotted in both diagrams. We see from the figure that 
only the stars with smallest mean orbital radii in Fig.\,\ref{fig_2}\,a 
closely follow our median curve at $[Fe/H]<-0.4$~dex. However at 
the largest distances, the [Mg/Fe]--[Fe/H] relation is almost linear 
(see the dashed line in Fig.\,\ref{fig_2}\,b).
In this case, only a small number of stars with sharply enhanced magnesium 
abundances are observed above the curve. At the same time, it can be 
noticed that the metallicity range for the bulk of the stars with 
$[Mg/Fe]<0.2$ is displaced from ($0.5<[Fe/H]<+0.3$) for the nearest 
stars to ($-0.7<[Fe/H]<+0.2$) for the farthest stars.  
The change in the behavior of the [Mg/Fe]--[Fe/H] relation with 
stellar orbital radius shows that the star formation rate closer to 
the Galactic center is higher than that on the periphery. Moreover it
seems that star formation within the solar circle of the Galaxy has 
never been interrupted, but only slowed down before the massive 
formation of thin-disk stars. In contrast, the first thin-disk stars at
great Galactocentric distances appeared only after the long phase of 
star formation delay. This follows from the presence of a distinct 
jump in the [Mg/Fe] ratio at metallicities $[Fe/H]<-0.4$~dex for stars
with large orbital radii. The larger [Mg/Fe] ratios at high 
metallicities in the stars within the solar circle suggest that the 
star formation rate remains there higher even at present. The clear 
deficit of stars with metallicities higher than the solar value there 
is also indicative of a lower star formation rate at great 
Galactocentric distances. 

Our interpretation was constructed on the suggestions that metallicity 
is good statistical age indicator in the thin disk. Indeed, lifetime 
of the thin disk is compared with the Galactic age, therefore the 
continuous process of the synthesis of chemical elements must lead 
during this period to a noticeable increase in the general abundance 
of heavy elements in the younger stars of subsystem.
As a result in the thin disk the well expressed trend of metallicity 
from the age must be observed. However, in spite of the large history 
of the study of this question, in regard to this there is no unanimous 
opinion. Twarog~(1980) was the first to derive the age--metallicity 
relation in the Galactic disk from F2--G2-stars and argued that it was 
unambiguous. However it was subsequently proven that the relation 
was by no means unambiguous and there was a significant spread in 
metallicity among the stars of any ages. This gave reason to suggest 
that there was no age--metallicity relation in the thin disk.
The purpose of this work is the thorough analysis of possible selective 
effects on the age--metallicity diagram with the attraction together 
with photometric data of the spectroscopic determinations of the iron 
and magnesium abundances for the main sequence stars over a wide range 
of spectral classes. For investigation we used 
Geneva-Copenhagen Surveu (Nordstr$\ddot e$m et al.~2004), where on the data
uvby$\beta$ photometry, and Hipparcos parallaxes were determined 
temperature, metallicity, distance, absolute magnitude, and ages for about 
14000~nearest F-G-K stars. The most probable ages was calculated on the 
base of Padova theoretical isochrones, using the sophisticated 
interpolation method, and Baysian computational techniques. 
After  removing of the binary stars, marked in the catalog, far 
evolving stars ($\delta M_V > 3^m$), and stars with uncertainly 
determined ages ($\varepsilon t >\pm 3$~Gyr), in the sample remained 
5540~supposedly single stars of thin disk. (Total average error in 
determination of age for the stars of the received sample comprised 
$\langle \varepsilon t\rangle = \pm 1.0$~Gyr.)
In order to get rid of the selective effects, connected with a 
difference in the depth of survey for the stars of different 
metallicity and temperature, we limited sample with the distance 
from the Sun equal to 70~pc, within limits of which our sample can 
be considered complete for temperature range ($5400 - 7200$)\,K$^\circ$. 
In finally formed thus sample remained 1890~stars of the thin disk.

\begin{figure}
\resizebox{\hsize}{!}
{\includegraphics{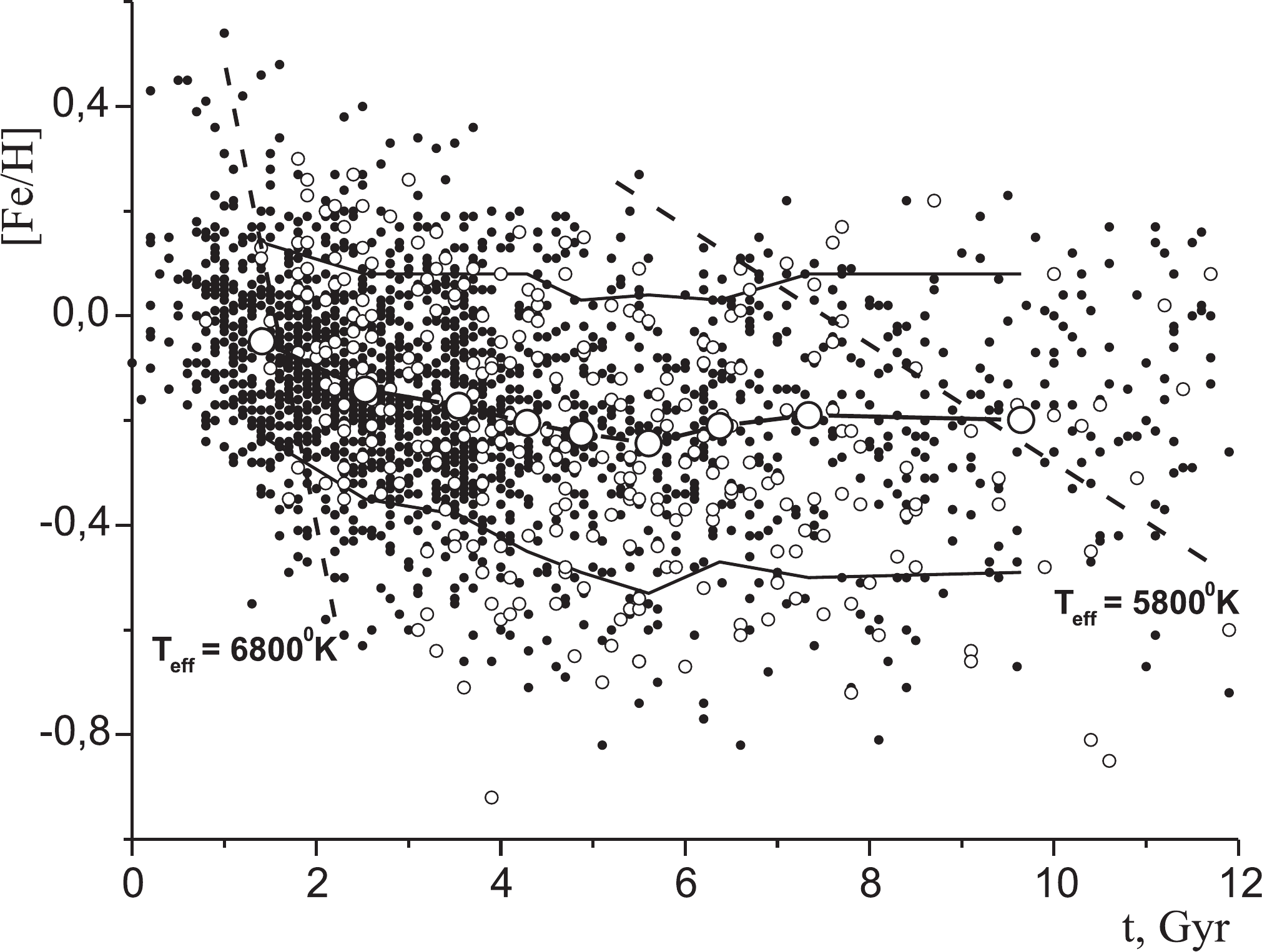}}
\caption{Age~-- metallicity diagram for the thin-disk stars from 
the catalog (Nordstr$\ddot e$m et al.~2004) with ($\varepsilon t>\pm3$~Gyr),
which lie nearer than 70~pc of the Sun. Large open circles connected 
by linear segments are the average values of the metallicity of stars in 
the narrow age ranges; broken lines - upper and lower ten-percent 
envelopes; inclined broken lines~-- theoretical isotherms for 
$T_{\textrm{eff}} = 6800$ and 5800~K; the small open circles~-- star 
with the spectroscopic determinations [Fe/H] from the catalog 
(Borkova, Marsakov,~2005).}
\label{fig_3}
\end{figure}

Figure\,\ref{fig_3} gives age~-- metallicity diagram for the 
thin-disk stars of our sample. The large opened circles in the 
figure put average values of the metallicity of stars in nine 
narrow ranges on age. Upper and lower ten-percent envelopes are 
also constructed. From the figure one can see that among the old 
stars of Galactic disk is observed large spread in values [Fe/H], 
whereas among the young stars the explicit scarcity of metal poor 
stars (empty left-hand lower corner on the diagram) is observed. 
(Nordstr$\ddot e$m et al.~2004) explained this by the effect of the 
limitation of their catalog from the high-temperature side on the 
index $(b - y)$. In this case, in their opinion, the difference 
at the ages of the metal rich and metal poor stars of the same 
temperature just also provides observed inclination of lower 
envelope on the diagram. Let us verify this statement.

\begin{figure}[t!]
\resizebox{\hsize}{!}
{\includegraphics{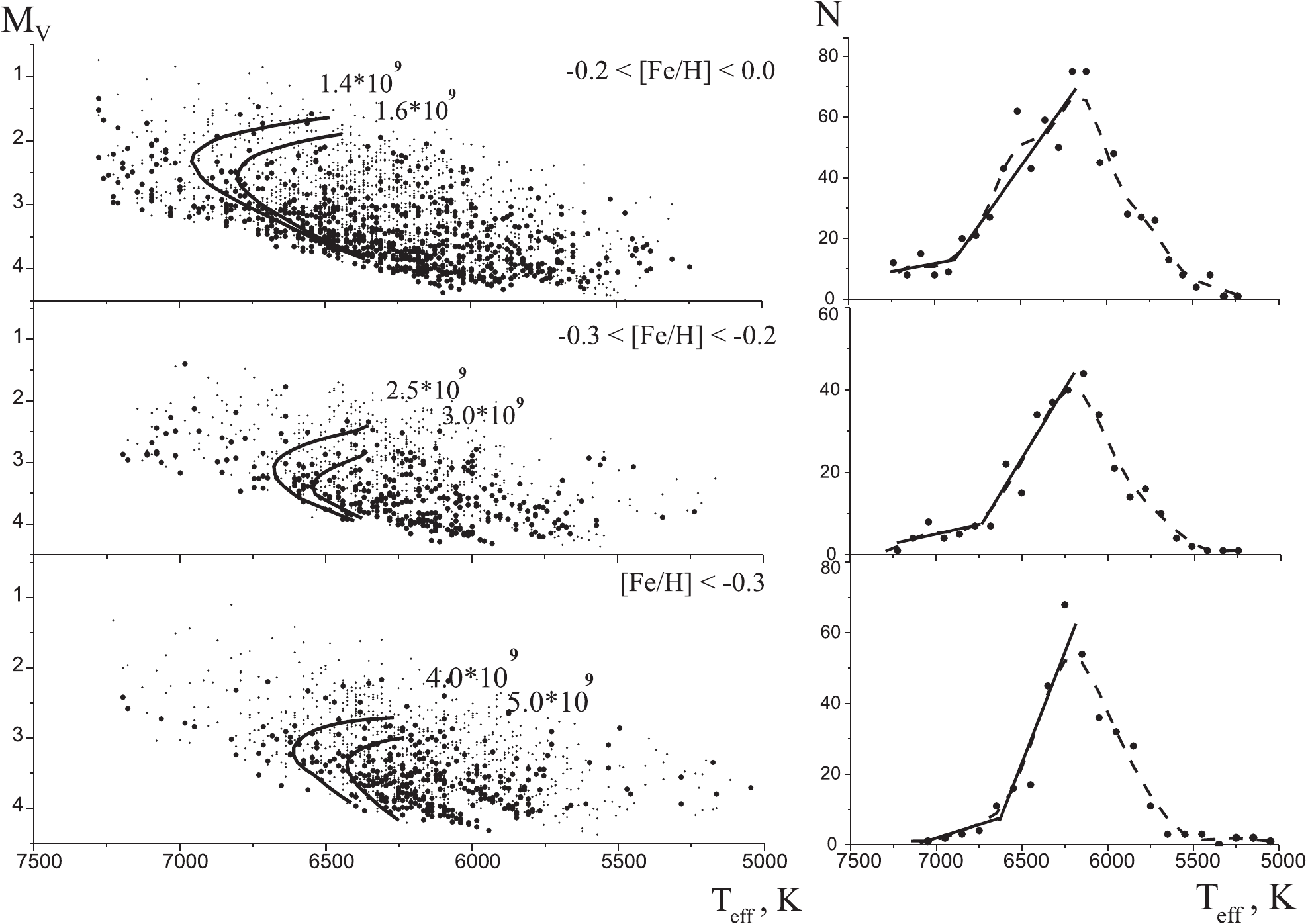}}
\caption{$T_{\textrm{eff}}- M_V$ diagrams for the metal-poor stars 
lying within 70~pc from the Sun into narrow ranges on the metallicity 
(to the left) and distribution the number of stars depending on 
temperature for the same diagrams (to the right). Broken dotted 
lines on the histograms - smoothed on three points trends with the 
sliding averages, linear segments emphasize the positions of sharp 
fractures on the left boundaries of distributions. On left panels 
are substituted theoretical isochrones from works (Demarque et al.,~2004), 
which correspond to the positions of fractures on the histograms 
(for comparison the isochrone of larger age are substituted more 
to the right them). The ages of isochrones are indicated.}
\label{fig_4}
\end{figure}

In Fig.\,\ref{fig_4} at the left the $T_{\textrm{eff}} - M_V$ 
diagrams for the stars lying within 70 pc from the Sun into three 
narrow ranges on the metallicity are given. On the right for the 
same diagrams the distributions of number of stars depending on 
temperature are constructed. Broken lines on the right panels 
represent trends, smoothed on three points, and linear segments 
schematically designate the behavior of the left boundaries of 
distributions. It is seen that all metal poor groups ($[Fe/H]<0.0$) 
demonstrate the sharp inflection envelope, when more to the right 
"inflection points" the number of stars abruptly increases. It is 
interesting that the temperature value of these points for all 
metal poor groups are far from the left edge of the diagrams. 
This means that the sharp scarcity of hotter (so also younger) 
stars in the metal poor groups is connected not to high-temperature 
limitation of sample but with existence of minimum age for the 
majority of the stars of given metallicity in the thin disk, i.\,e., 
with existence of "turnoff points" in the metal poor stars of field. 
On the left panels of figure the theoretical isochrones, passing 
through those isolated on the right panels "inflection points",  
are carried out according to the data of work (Demarkque et al.,~2004). 
For the comparison are more to the right are everywhere substituted 
the isochrones of larger age

The effect of the limitation of sample from the high-temperature 
side, which it is discussed in the work ( Nordstr$\ddot e$m et al.~2004), 
also somewhat distorts real age~-- metallicity diagram. In 
figure\,\ref{fig_3} the dash inclined line is conducted on the 
basis of the theoretical isochrones for $Ò_{\textrm{eff}} = 5800$~K. 
This line corresponds to boundary, more to the right of which 
stars, hotter this temperature, cannot be~-- they have left from 
the main sequence already. (Let us note that precisely the 
limitation of sample from the low-temperature side with approximately 
Solar value of temperature (see the dash inclined line in the upper 
right-hand corner in the diagram) led to the exception of the oldest 
metal rich stars from the sample (Twarog,~1980), which caused his 
conclusion about a monotonic increase of the metallicity with the 
age in the thin disk.)

However, as can be seen from the diagrams, not this line, which 
intercepts an entirely small quantity of very young metal poor 
stars, but relative numbers of stars of different metallicity with 
the identical age determine the variation of average metallicity 
on the age. Note that lower envelope in Fig.\,\ref{fig_3} in the 
range $(1<t<4)$~Gyr practically coincides with left envelope of 
diagram. This means, that lower envelope reflects the variation 
of the "turnoff points" position of the  stars of this metallicity 
on the age.  As a result we see that "turnoff points" of metal 
poor stars are located in the middle of the temperature range, 
occupied by the stars of the sample, i.\,e., they are not 
connected with the action boundary selective effect.

To the distortion of the ages of stars can lead also the effect 
of their unresolved binary. Actually, the luminosities of the 
unresolved close binaries, calculated from the trigonometric 
parallaxes, are more than true (Suchkov,~2000). As a result the 
ages of stars not yet reached their turnoff point will be 
overestimated, whereas higher than it are located ever younger 
stars at an identical temperature. We isolated in our sample 
candidates into the dual according to the criterion, proposed 
in latter work. Such proved to be about 20\,\%, but constructed 
for the remained single stars age~-- metallicity diagram 
demonstrated that its form practically did not change.

\begin{figure}
\resizebox{\hsize}{!}
{\includegraphics{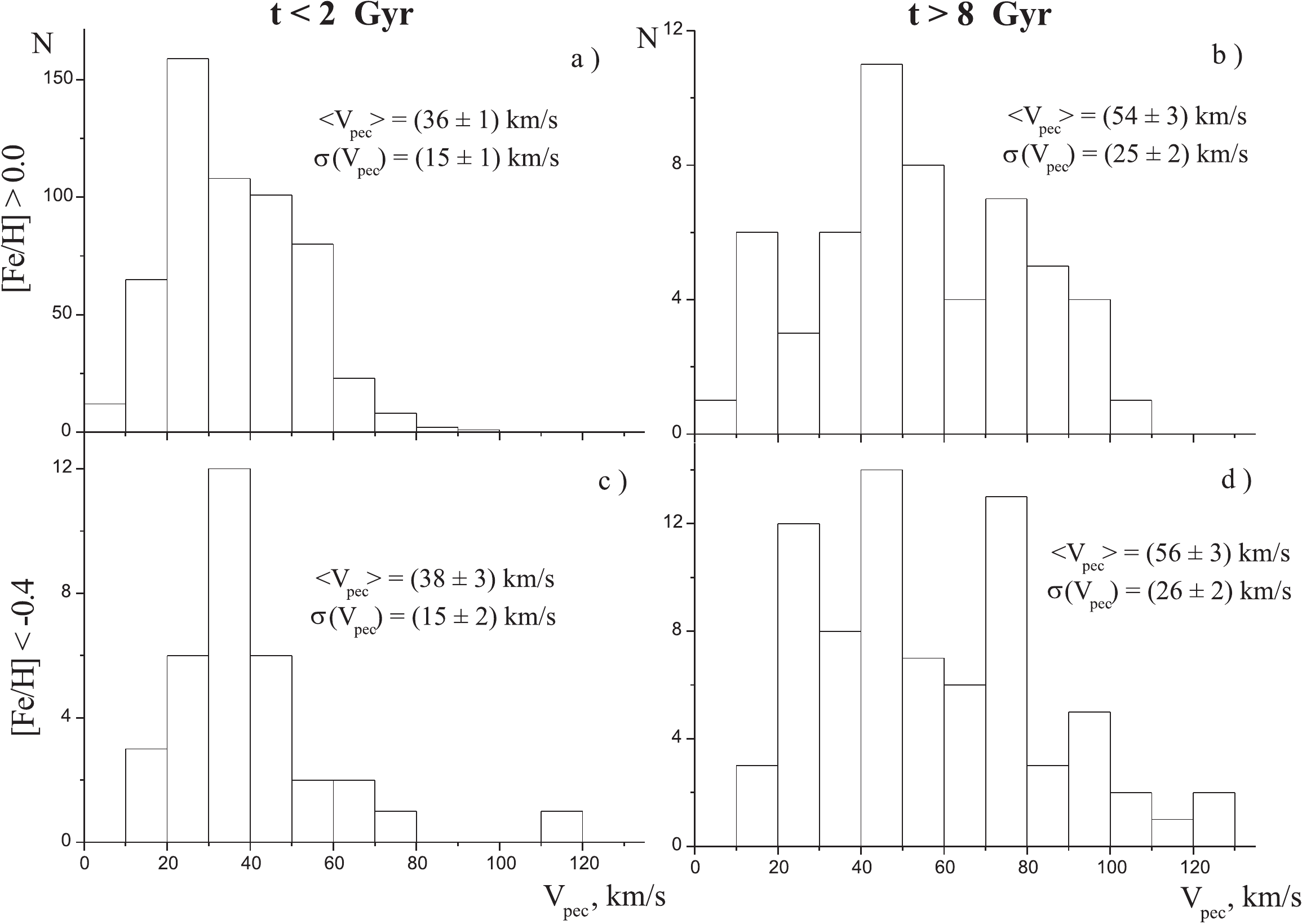}}
\caption{Distributions on peculiar velocities of metal-rich and 
metal-poor stars of different age. The corresponding values of 
average velocities and their dispersions are indicated on the 
panels. It is seen the coincidence of histogram of similar age.}
\label{fig_5}
\end{figure}

Despite the fact that the position of a sufficiently large 
quantity of stars with the solar metallicity on the 
Hertzsprung - Russel diagram indicates their large ages, 
in the work (Pont, Eyer,~2004) existence of old metallic stars 
undergoes doubt. 

Therefore let us verify, actually whether some metal rich stars 
are actually old? It is possible to do on the independent 
statistical indicator of age~-- kinematics. For this let us compare 
peculiar velocities distributions of metal rich ($[Fe/H]>0.0$) 
and metal poor ($[Fe/H]<-0.4$) thin-disk stars, preliminarily 
isolate among them very young ($t<2$~Gyr) and very old ($t>8$~Gyr) 
star. The indicated histograms are given in Fig.~\ref{fig_5}. 
Comparison shows that the distributions of the stellar velocity 
of identical age, but different metallicity, are very similar. 
In this case old stars demonstrate one and a half times higher 
in both the average values and dispersions of peculiar velocities 
than young stars (see inscription on the appropriate panels). 
This behavior makes it possible to assert that the some metal 
rich stars have actually very large age.

\begin{figure}[t!]
\resizebox{\hsize}{!}
{\includegraphics{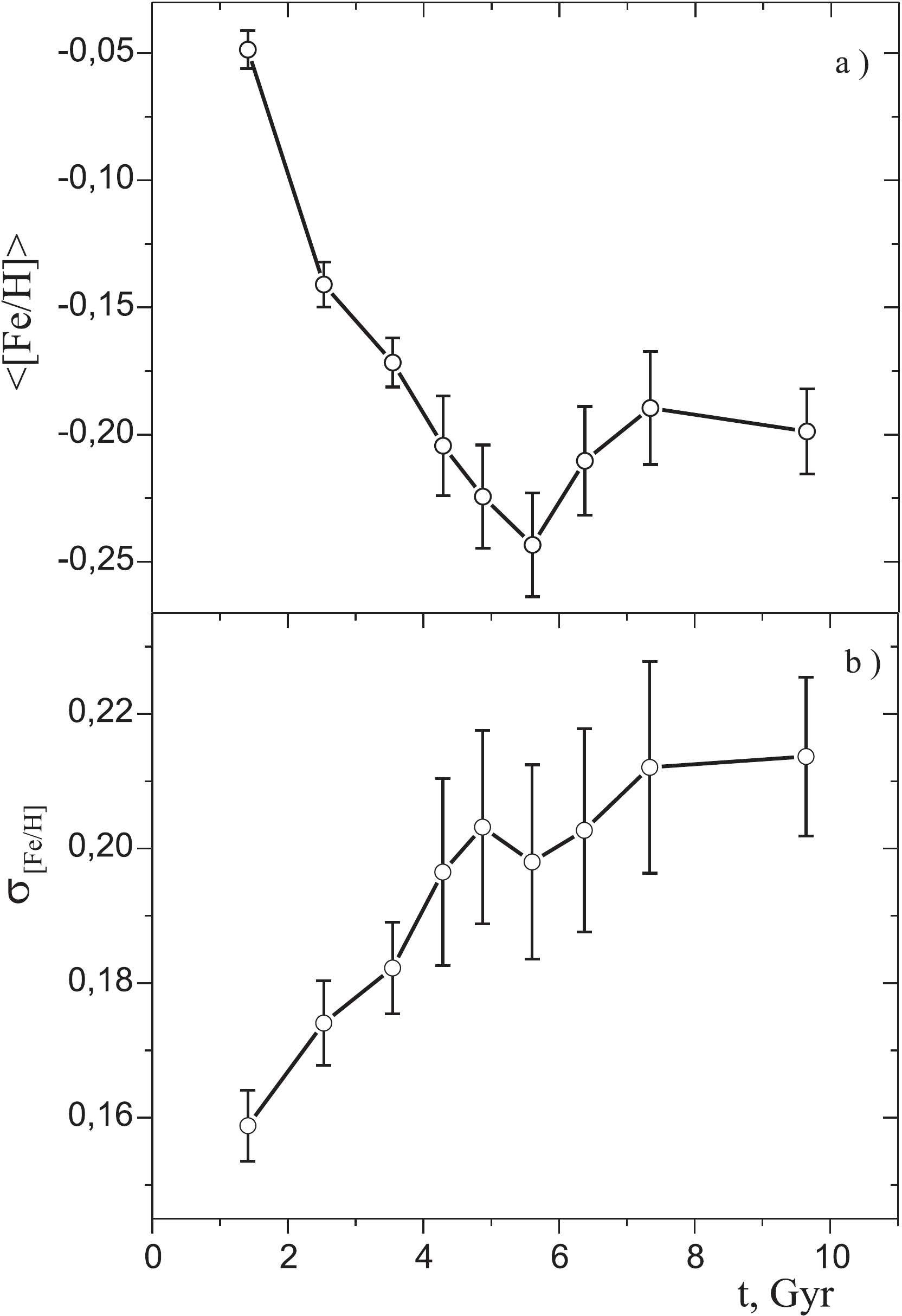}}
\caption{Dependences of average value (a) and dispersion of 
metallicity (b) from the age for the stars of thin disk in the 
70~pc of the Sun from the catalog (Nordstr$\ddot e$m et al.~2004). 
Bars are error in determination of the corresponding values.}
\label{fig_6}
\end{figure}

We assume that none of the selective effects the known to us 
cannot substantially distort the common form of age~-- metallicity 
diagram  in Fig.\,\ref{fig_3} for the sample of F-G stars in the 
70~pc of the Sun. Therefore let us trace the behavior of the 
dependences of average value and dispersion of metallicity from 
age, constructed on the basis of these data. From Fig.\,\ref{fig_6}\,a, 
where the corresponding values were calculated in 9 narrow bins from 
the age, can be seen that at first average metallicity noticeably 
decreases with an increase in the age, and after $\approx5$~Gyr 
it remains practically constant and equal to 
$\langle [Fe/H]\rangle =-0.22$. (Local minimum in the environment of 
5~Gyr is most likely connected with the special features of the 
age determination procedure in the work (Nordstr$\ddot e$m et al.~2004).) 
Dispersion of metallicity, as can be seen from Fig.\,\ref{fig_6}\,b, 
always monotonically increases from 0.16 to 0.21; however, after 
$\approx 5$~Gyr this increase significantly slows down.

\begin{figure}[t!]
\resizebox{\hsize}{!}
{\includegraphics{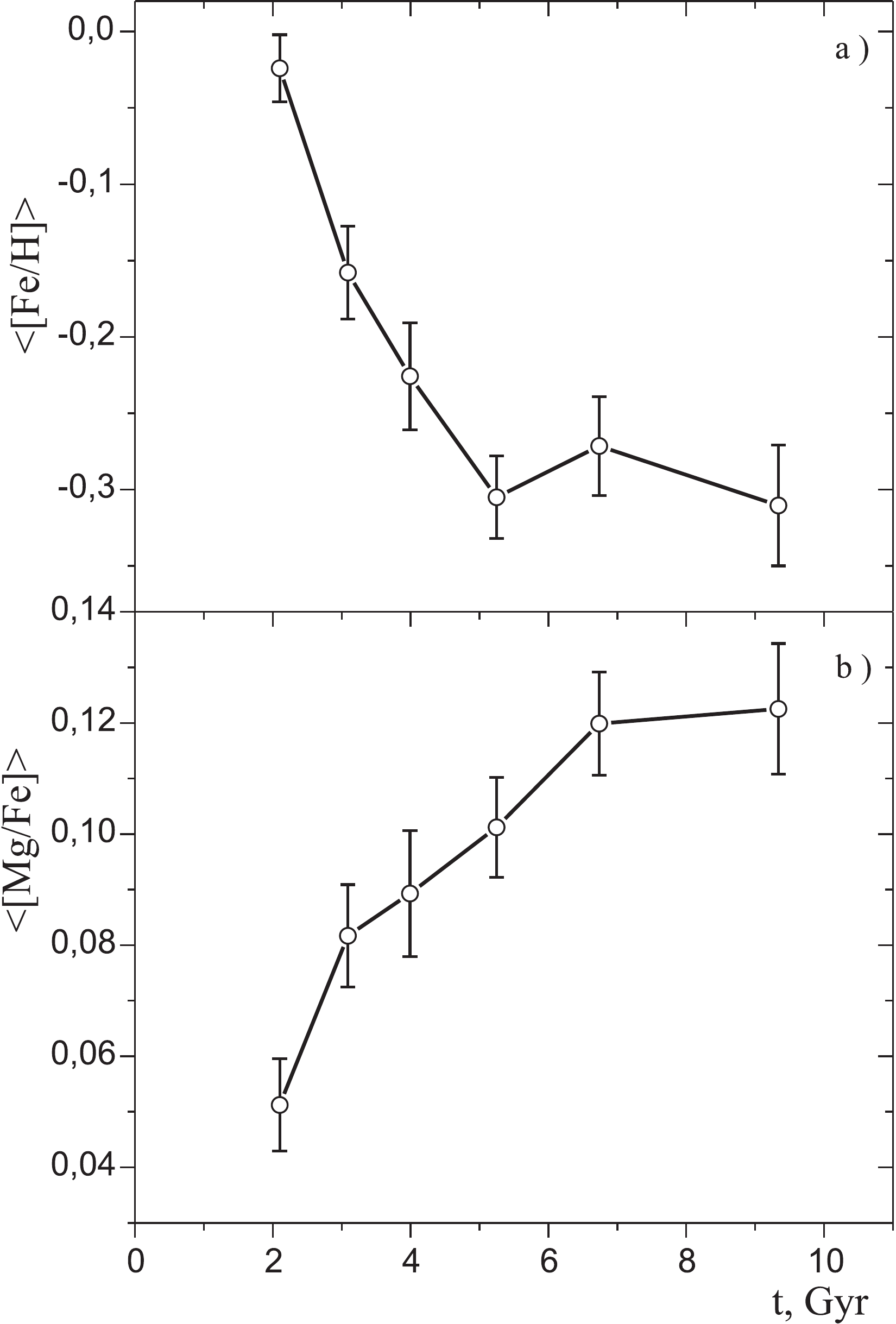}}
\caption{Dependences of average values $\langle [Fe/H]\rangle$ (a) and 
$\langle [Mg/Fe]\rangle$ (b) from the age for the thin-disk stars with 
the spectroscopically specific abundances of iron and magnesium from 
(Borkova, Marsakov,~2005). Bars are error in determination of the 
corresponding values.}
\label{fig_7}
\end{figure}

Photometric metallicity in some stars can be distorted by the 
disregarded systematic effects; therefore it is necessary to 
investigate age~-- metallicity dependence, also on the stars of 
our catalogue with the spectroscopic determinations of [Fe/H] 
(see the small open circles in Fig.~\ref{fig_3}). Let us note that 
in this sample the unresolved binaries deliberately be absent~-- 
otherwise they with the great probability would be known as 
spectrally binaries. Let us recall that since the number of 
stars here is not very great, we left all stars of thin disk 
in the sample. It is seen that on this diagram be absent very 
metal rich ($[Fe/H]>0.3$) star~-- the apparently photometric 
"super-metallicity" of these stars actually is explained as 
an artifact of the deredding procedure for distant stars as 
this predicted in the work (Nordstr$\ddot e$m et al. 2004). 
In other respects the form of diagram practically did not 
change. From  Fig.~\ref{fig_7}\,a we see that age~-- metallicity 
relation, constracted on the stars of catalog with the 
spectroscopic determinations of [Fe/H], demonstrates the 
same behavior.

For understanding of the reason for such complex behavior of 
age~-- metallicity dependence it is important to trace the 
dependence of the relative content of magnesium on the age 
among the disk stars. Age -- magnesium abundance diagram itself 
according to the data of our catalog is given in Fig.\,\ref{fig_8}\,[b] 
in the work (Marsakov, Borkova,~2006), here in Fig.~\ref{fig_7}\,b 
is represented only the variation of the relation $\langle [Mg/Fe]\rangle$ 
from the age. From the Fig.\,\ref{fig_7}\,b one can also see that, 
as in the case the connection between the age and the metallicity, 
bend is observed in the middle of the dependence between relative 
magnesium abundance and age. Bend evidences that the relative 
magnesium abundance in the thin-disk stars of being in the subsystem 
formation initial stages was sufficiently high 
($\langle [Mg/Fe]\rangle \approx 0.10$). About 5~billion years 
ago it began to sharply decrease with the approximation to the 
present time. Thus, taking into account the uncertainty of the 
estimations of average values, we can assert that an increase 
in the average metallicity and the decrease of the average 
relative magnesium abundance in the stars of thin disk began 
simultaneously. 

Thus components, confidently revealing on both $t - [Fe/H]$ and 
$t - [Mg/Fe]$ diagrams, testify that into the first several billion 
years of the formation of the thin-disk subsystem interstellar 
medium in it was, on average,  sufficiently rich in heavy elements 
($\langle [Fe/H]\rangle \approx -0.20$) and is badly mixed 
($\sigma_{[Fe/H]}\approx 0.21$), and the average relative 
abundance of magnesium was comparatively high 
($\langle [Mg/Fe]\rangle \approx 0.10$). Approximately 5~billion 
years ago the average metallicity began to increase,
and the dispersion of metallicity and the relative magnesium 
abundance~-- to decrease. This occurred as a result sharply 
increased rate of star formation and making more active of the 
processes of mixing to the interstellar medium. By the possible 
reason for this could be interaction of the Galaxy with the 
completely massive satellite galaxy.

Thus, not all stars of the thin disk, which are at present 
located in the Solar neighborhood, were formed from the matter, 
which experienced united chemical evolution. We suppose that the
difference in the star formation rate at the different 
galactocentric distances and sporadic fall out to the disk of 
gas from the exteriors of the Galaxy led to the ambiguity of 
dependence between the age and the metallicity in the so 
long-life subsystem.

The complete description of the first part of this work was 
published in (Marsakov, Borkova,~2006), but the second part 
will be published latter.

{\it Acknowledgements.} This work was supported in part by the 
Federal Agency for Education (projects RNP~2.1.1.3483 and 
RNP~2.2.3.1.3950) and by the Southern Federal University (project 
K07T~-- 125).
\\[3mm]
\indent

{\bf References\\[2mm]}
1.~T.~Bensby, S.~Feldsing and I.~Lundstrem, Astron. Astrophys. {\bf 410}, 
527 (2003)\\
2.~T.V.~Borkova and V.A.~Marsakov, Astron.Zh. {\bf 82}, 453 (2005)
[Astron.Rep.49 405 (2005)].\\
3.~P.~Demarque, J-H.~Woo, Y-C.~Kim, S.K.~Yi, Astron.Astrophys.Suppl.Ser. 
{\bf 155}. 667 (2004).\\
4.~B.~Nordstr$\ddot e$m, M.~Mayor, J.~Andersen et~al., 
Astron. Astrophys. {\bf 418}, 989 (2004)\\
5.~V.A.~Marsakov and T.V.~Borkova. Pis'ma Astron.Zh. {\bf 32}, 419 (2006)\\
6.~F.~Pont, L.~Eyer), Mon.Not.Roy.Astron.Soc., {\bf 351}, 487 (2004).\\
7.~A.A.~Suchkov, Astroph.J, {\bf 535}. L 107\\
8.~B.M.~Tinsley, Astrophys. J. {\bf 229}, 1046 (1979)\\
9.~B.A.~Twarog, Astrophys. J. {\bf 242}, 242 (1980)
\\

\end{document}